\documentclass[iop]{emulateapj}
\usepackage{apjfonts}
\usepackage{hyperref}
\hypersetup{
    colorlinks=true,
    linkcolor=blue,
    filecolor=magenta,      
    urlcolor=blue,
    citecolor=blue
}
\usepackage{color}
\usepackage{textcomp}
\usepackage{amsmath}
\shorttitle{Observations of Reconnection Flows in a Flare on the Solar Disk}
\shortauthors{Wang et al.}

\begin{document}

\title{Observations of Reconnection Flows in a Flare on the Solar Disk}

\author{Juntao Wang, P. J. A. Sim\~{o}es, N. L. S. Jeffrey, L. Fletcher, P. J. Wright, and I. G. Hannah}
\affil{SUPA, School of Physics and Astronomy, University of Glasgow, Glasgow G12 8QQ, UK}
\email{j.wang.4@research.gla.ac.uk}

\begin{abstract}
Magnetic reconnection is a well-accepted part of the theory of solar eruptive events, though the evidence is still circumstantial. Intrinsic to the reconnection picture of a solar eruptive event, particularly in the standard model for two-ribbon flares (``CSHKP'' model), are an advective flow of magnetized plasma into the reconnection region, expansion of field above the reconnection region as a flux rope erupts, retraction of heated post-reconnection loops, and downflows of cooling plasma along those loops. We report on a unique set of SDO/AIA imaging and Hinode/EIS spectroscopic observations of the disk flare SOL2016-03-23T03:54 in which all four flows are present simultaneously. This includes spectroscopic evidence for a plasma upflow in association with large-scale expanding closed inflow field. The reconnection inflows are symmetric, and consistent with fast reconnection, and the post-reconnection loops show a clear cooling and deceleration as they retract. Observations of coronal reconnection flows are still rare, and most events are observed at the solar limb, obscured by complex foregrounds, making their relationship to the flare ribbons, cusp field and arcades formed in the lower atmosphere difficult to interpret. The disk location and favorable perspective of this event have removed these ambiguities giving a clear picture of the reconnection dynamics.
\end{abstract}

\keywords{Sun: corona --- Sun: coronal mass ejections (CMEs) --- Sun: filaments, prominences --- Sun: flares --- Sun: solar wind --- Sun: UV radiation}

\section{Introduction}\label{intro}
Solar eruptive events, flares and associated coronal mass ejections (CMEs), are attributed to the liberation of free magnetic energy stored in the corona, possibly due to magnetohydrodynamic  (MHD) instabilities and magnetic reconnection. The ``CSHKP'' model is the standard  2D framework for two-ribbon flares \citep{car1964,stu1966,hir1974,kop1976}, and predicts several different flows in the flare corona. There is an inflow of plasma and magnetic field towards a diffusion region where reconnection occurs, 
and an outflow from this region of newly-reconnected field retracting due to magnetic tension. Both flows are (roughly) perpendicular to the magnetic field direction. There is cooling, condensing material flowing along post-reconnection loops down towards the solar surface. The flare or eruption may influence the ubiquitous upflows at the edge of the active region (AR). In this paper we show that a plasma upflow parallel to the inflow field may also happen due to a ``depressurization'' process as the field erupts.

Evidence for reconnection inflows has been reported in a handful of flares, mainly at the solar limb. \citet{yok2001} reported the first clear extreme ultraviolet (EUV) inflow following an eruption, with a bright cusp  --  another ingredient in the ``CSHKP'' model -- seen underneath in soft X-rays (SXR). \citet{nar2006} found a further 6 limb inflow events in nearly 5 years of Extreme-ultraviolet Imaging Telescope (EIT) observations. A bright, elongated structure in the inflow convergence region was claimed by \citet{lin2005} to be a current sheet, and the features flowing up along it to be reconnection outflows. A few more inflows have been reported using observations from the Atmospheric Imaging Assembly \citep[AIA;][]{lem2012} onboard the Solar Dynamics Observatory \citep[SDO;][]{pes2012}. \citet{sav2012} studied an inflow with speed up to $\rm\sim300~km~s^{-1}$ in an impulsive flare, while other reports, usually of long duration events (LDEs) have speeds below $\rm\sim100~km~s^{-1}$. \citet{sunj2015} reported groups of inflowing ``threads'' with plasma heating where they make contact, but without a clear hot cusp. In 3 different flares, \citet{su2013}, \citet{yang2015}, and \citet{zhu2016} observed a reconnection inflow with two sets of closed loops approaching each other - a different geometry from the standard  model.

Reconnection outflows -- the retraction of post-reconnection magnetic loops -- have occasionally been reported in SXR limb flares \citep{for1996,ree2008}, but  EUV is better at picking out retracting structures. \cite{liu2013} detected many individual retracting loops in AIA 131 {\AA}~ observations of a limb flare, with speeds from tens to hundreds of $\rm~km~s^{-1}$. \cite{ima2013} combined AIA and Hinode EUV Imaging Spectrometer \citep[EIS;][]{cul2007} observations to infer that the hot reconnected loops $\sim30$ MK could shrink at a speed above $500 \rm~km~s^{-1}$. Supra-arcade downflows, the dark voids in EUV and SXR observations appearing high in the corona and traveling down at tens to hundreds of $\rm~km~s^{-1}$, are interpreted as the cross-sections of underdense, retracting post-reconnection loops, or the `wakes' left as they descend \citep[e.g.,][]{mck1999}. Plasma draining in flare loops as reconnection downflows has also been observed \citep[e.g.,][]{sav2012}. EIS spectroscopic observations shows that the draining speed along AR loops at quiescent stage (when there is no flare or eruption) is around tens of $\rm~km~s^{-1}$ \citep{del2008,syn2012}.

The inflow Alfv\'{e}n Mach number defining the reconnection rate for these events is estimated at $\sim10^{-1}-10^{-3}$ in the fast reconnection regime \citep[the slow Sweet-Parker rate is $\sim10^{-4}-10^{-6}$ for typical coronal conditions;][]{asc2005}. But a good estimate of the reconnection rate requires knowledge of the coronal magnetic field strength, which is difficult to obtain in the limb events stated above. Their position also makes the relationship between the cusp, loops and footpoints hard to ascertain as the footpoints are usually obscured by the solar limb or complex foreground structures.

We report here on a long-lasting reconnection event near the disk center, focusing on its flow processes and magnetic reconnection rate. \citet{li2017} studied this event using SDO/AIA, demonstrating the relationship between the erupting flux rope and magnetic reconnection, and the transition from 3D to 2D reconnection.  The event's location and quasi-2D geometry in the late phase permit a good estimate of the coronal Alfv\'{e}n speed and reconnection rate. It exhibits the norms of the standard ``CSHKP'' model, with a well-formed cusp underneath inflow threads which can be mapped well to their lower-atmosphere counterparts. The field below the cusp contracts and cools (though the brightest portion rises). We also find spectroscopic evidence for a new kind of plasma upflows associated with the expanding but closed inflow field during a flare, distinct from the common plasma upflows at the AR boundary at the quiescent stage that have been reported by previous authors.

\section{Observations and Analyses}\label{obs}
\subsection{Instruments and Data Reduction}\label{ins}
SOL2016-03-23T03:54 was a Geostationary Operational Environmental Satellite (GOES) class C1.1 flare in AR NOAA 12524 (N15W16). We study it from $\sim$ 01:00 UT to $\sim$ 07:00 UT. The SDO/AIA and Helioseismic and Magnetic Imager (HMI; \citeauthor{scho2012} \citeyear{scho2012}) provide EUV images and photospheric magnetograms, respectively, which have been processed using standard software \citep{boe2012} and rotated to 01:00 UT. The EIS on Hinode observes the AR in a slow raster from 04:01:50 UT to 05:02:42 UT with a 1" slit moving around every minute from solar west to east over a  field-of-view 119.8"$\times$512.0". Line-of-sight velocities are obtained from Fe XII and Fe XIII lines, which are intense and also visible outside the active region, for estimating a reliable rest wavelength. Standard EIS data reduction procedures were used, and the spectral lines were fitted with single Gaussians. The rest wavelength was extracted from a quiet Sun region X$\sim(-24'', 85'')$ and Y$\sim(157'', 207'')$ (excluding missing values along a vertical data gap at X$\sim13''$) free of AR emission. The upper-limit uncertainty is $\sim5~\rm km~ s^{-1}$ for both Fe XII 195.12 {\AA} and Fe XIII 202.04 {\AA}, and Fe XVI 262.98 {\AA} has an upper-limit uncertainty $\sim9~\rm km~ s^{-1}$. The alignment between AIA and EIS is conducted by eye, and also takes Fe IX 197.86 {\AA} into account (but Fe IX intensity is too low for reliable Doppler velocity diagnostics). Fe IX is aligned with 171 {\AA}, Fe XII with 193 {\AA}, Fe XIII with 211 {\AA}, and Fe XVI with 335 {\AA}, as their characteristic temperatures are comparable seperately. The accuracy of the alignment is $\sim1-2$ arcsecs.

\subsection{Evolution of the Flare}
Figure~\ref{evocut} shows the overall evolution of the flare. Before the flare (Figure~\ref{evocut}(a)) a large arcade of loops in 171 {\AA} envelopes a dark void underneath, possibly a flux rope \citep{li2017}. Between the arcade footpoints a filament can vaguely be seen  (Figure~\ref{evocut}(b) and (e) show the filament more clearly). In Figure~\ref{evocut}(b), the two ends of the filament suddenly brighten (microflare), accompanied by a small ejection to the north. This may show the destabilization of the hosted flux rope, leading to the subsequent arcade eruption in Figure~\ref{evocut}(c). As the arcade erupts, its legs converge, forming a dark cusp underneath in 171 {\AA}, shown in Figure~\ref{evocut}(d). The flare ensues with a bright cusp in 131 {\AA} (red) inside the dark cusp in 171 {\AA}. Then two ribbons sweep across the footpoints of the bright cusp and separate away from the filament, seen in 304 {\AA} in Figure~\ref{evocut}(e). Figure~\ref{evocut}(f) shows the post-flare state with flaring loops appearing in 171 {\AA}. The main evolution from Figure~\ref{evocut}(a), (c), (d) and (f) reveals that the correspondence between the pre-flare arcade, the erupting arcade, the bright cusp and the flaring loops is well established in terms of their footpoint locations, indicated by the two magenta circles. Figures~\ref{allts}(a)-(c) show the timeslices corresponding to cuts 1-3 in Figure~\ref{evocut}, respectively. A lightcurve in 304 {\AA} for the microflare in Figure~\ref{evocut}(b) is added in Figure~\ref{allts}(c), and the GOES SXR lightcurves in Figure~\ref{allts}(d). The vertical dotted line ``A'' indicates the timing of the microflare and the arcade eruption, and the line ``B'' the timing of the inflow and the flare. Different flows are discussed in the following paragraphs.

\subsection{Flows in the Flare}\label{allflows}
\begin{figure*}
\hspace{-1.8cm}\includegraphics[scale=1.2]{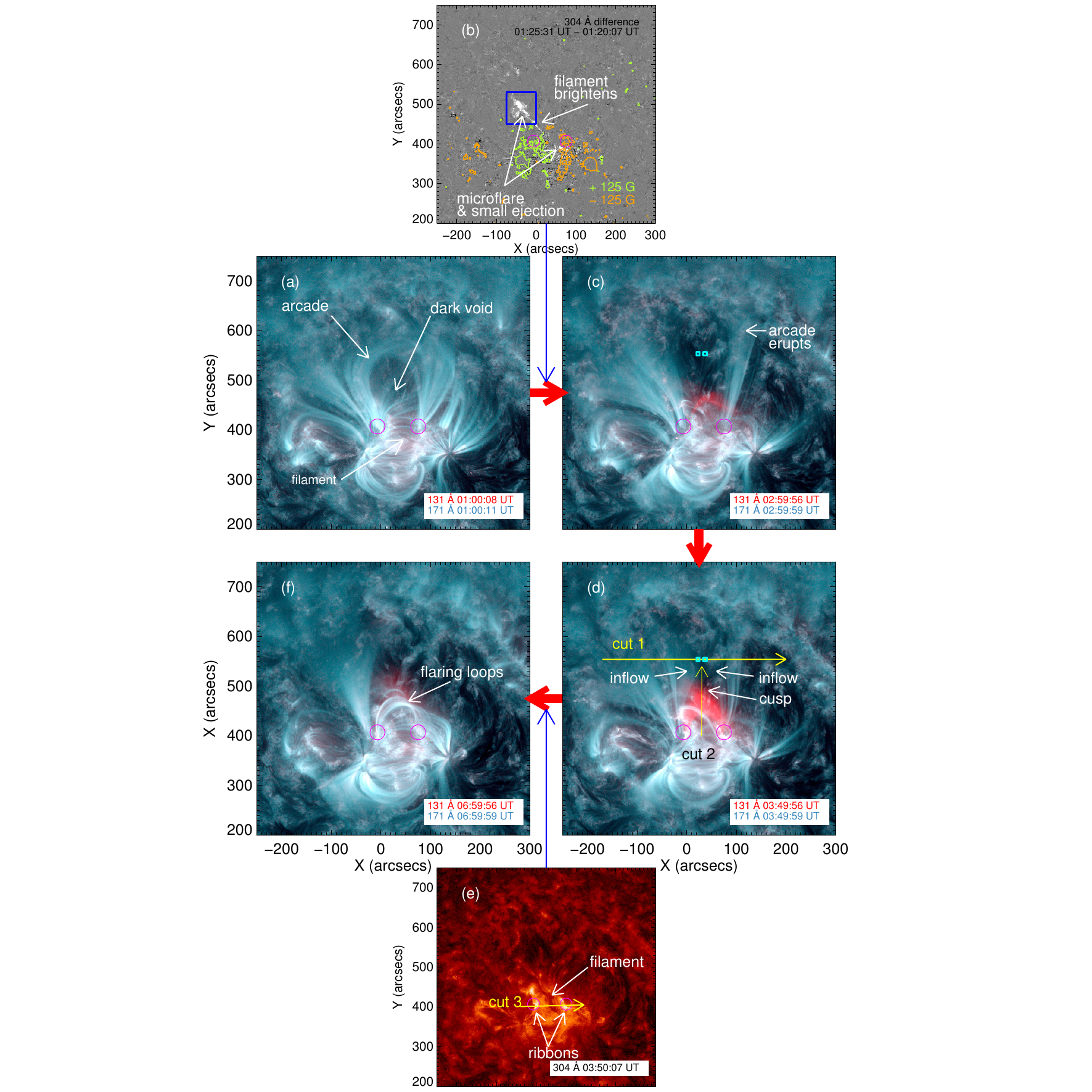}
\caption{\label{evocut}Evolution of the flare. (a), (c), (d) and (f) show the main evolution sequence in composite AIA 131 {\AA} and 171 {\AA} images. (b) The microflare and small ejection in the 304 {\AA} difference image just before the arcade eruption. The HMI magnetogram contours at $\pm125$ G are overlaid. The blue rectangle is used for the lightcurve in Figure~\ref{allts}(c). (e) The ribbon separation in 304 {\AA}. The magenta circles in each image show the relevant footpoint locations. Cuts 1-3 are used for timeslices in Figures~\ref{allts}(a)-(c), respectively. The two cyan boxes in (c) and (d) are for DEM analysis in Section~\ref{dem}. An animation of this figure is available.}
\end{figure*}

\begin{figure*}
\includegraphics[scale=0.9]{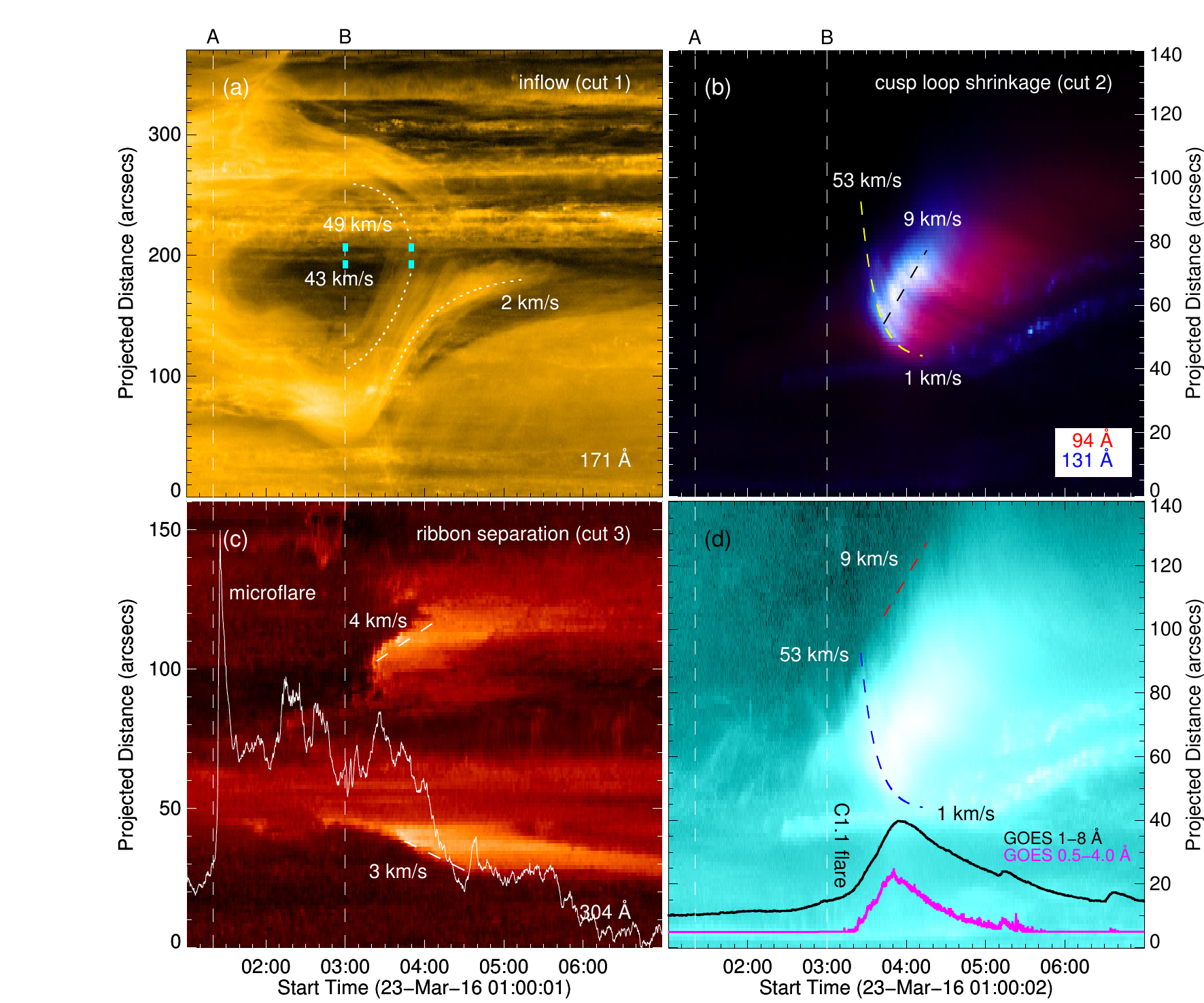}
\caption{\label{allts}(a) Timeslice of Cut 1 for the evolution of the inflow threads. The two cyan boxes at 03:00 UT and also 03:50 UT show the positions used for DEM analysis in Section~\ref{dem}, as in Figures~\ref{evocut}(c) and (d). The speeds given are for the final times of the fit curves. (b) Timeslice of cut 2 combines 94 {\AA} and 131 {\AA} on a linear intensity scale, showing the evolution of the bright cusp. Its corresponding image on a logarithmic intensity scale in 131 {\AA} is plotted in (d). The yellow dashed fit curve is the same as the blue one in (d), and the black dashed fit curve is the same as the red one in (d) but moved downwards to match the brightest portion. (c) Timeslice of cut 3  for the evolution of the ribbons. The lightcurve in 304 {\AA} of the microflare indicated in Figure~\ref{evocut}(b) is overlaid. (d) GOES SXR lightcurves overlaid on the timeslice image of cut 2 in 131 {\AA} on a logarithmic intensity scale for reference. The dotted line ``A'' denotes the timings of the arcade eruption and the microflare, and ``B'' the timings of the inflow and the C1.1 flare.}
\end{figure*}

\paragraph{Reconnection Inflows}
Figure~\ref{allts}(a) shows the evolution along cut 1 through the flare cusp region. Before the flare the threads forming the arcade legs separate as the flux rope erupts. The threads then accelerate towards the (presumed) central diffusion region, approaching with projected speeds of tens of $\rm km~s^{-1}$, similar on either side. These are fitted with exponential equations by picking a few points along specific inflow features and extrapolated to the diffusion regions indicated by the cyan boxes. The speeds at the final times of the fit curves are larger than that in \cite{li2017}, because we choose a cut with higher altitude than theirs, closer to the reconnection site at 03:50 UT in Figure~\ref{evocut}(d), in order to account for the progressively higher up reconnection site. Accelerated inflows were also found by \citet{sunj2015} and \citet{zhu2016}. After the GOES peak, the western leg gradually fades, while the flow of the outer threads of the eastern leg starts to decelerate towards the central region, reducing to a few $\rm km~s^{-1}$. Figure~\ref{allts}(c) shows the corresponding ribbon separation, also with similar speed on each side.

\paragraph{Reconnection Outflows}
The post-reconnection outflow is manifested as contraction of the loops underneath the cusp, visible as bright and dark striations in the stackplot (Figure~\ref{allts}(b)) of superposed 131 {\AA} and 94 {\AA} slices, on a linear intensity scale, along cut 2 vertically down through the cusp loops (also can be seen in the reference image of Figure~\ref{allts}(d) on a logarithmic intensity scale in 131 {\AA}). The yellow dashed line in Figure~\ref{allts}(b) shows the looptop in the cusp declining in altitude with time, illustrating the contraction of the cusp loops. The contraction decelerates with time, while the loops also cool down from 131 {\AA} ($\sim$10 MK) to 94 {\AA} ($\sim$6.8 MK). This is not well observed in the past to our knowledge. Meanwhile the brightest portion of the cusp rises, as expected if the reconnection site progressively moves upwards. We note the qualitative similarity between the observed trajectories of the contracting loops and those calculated by \cite{lin2004} for a 2D reconnecting current sheet model.

\begin{deluxetable*}{ccccccccccccc}
\tabletypesize{\small}
\tablewidth{0pt}
 \tablecaption{Magnetic Reconnection Parameters.\label{para}}
 \tablehead{
 \colhead{Region}  & \colhead{$V_{patt}$\tablenotemark{a}} & \colhead{$V_{xp}$\tablenotemark{b}} & \colhead{$\theta$\tablenotemark{c}} & \colhead{$V_{in}$\tablenotemark{d}} & \colhead{$V_{foot}$\tablenotemark{e}} & \colhead{$B_{foot}$\tablenotemark{f}} & \colhead{$B_{in}$\tablenotemark{g}} & \colhead{$\mathrm{EM}_{in}$\tablenotemark{h}} & \colhead{$L$\tablenotemark{i}} & \colhead{$n_e$\tablenotemark{j}} & \colhead{$V_{A}$\tablenotemark{k}} & \colhead{$M_{A}$\tablenotemark{l}} \\
 \colhead{}  & \colhead{($\rm km~s^{-1}$)} & \colhead{($\rm km~s^{-1}$)} & \colhead{degree} & \colhead{($\rm km~s^{-1}$)} & \colhead{($\rm km~s^{-1}$)} & \colhead{(G)} & \colhead{(G)} & \colhead{($\rm 10^{25}~cm^{-5}$)} & \colhead{$\rm arcsec$} & \colhead{($\rm 10^{8}~cm^{-3}$)} & \colhead{($\rm km~s^{-1}$)} & \colhead{}}
 \startdata
Eastern  & 43 & 9 & 27 & 38 & 3 & 131 & 10 & 4.8 & 15 & 2.1 & 1371 & 0.03\\
Western  & 49 & 9 & 27 & 44 & 4 & -125 & 11 & 4.3 & 15 & 2.0 & 1551 & 0.03
\enddata
\tablenotetext{a}{obtained from Figure~\ref{allts}(a).}
\tablenotetext{b}{estimated from the rising speed of the bright cusp in 131 {\AA} in Figure~\ref{allts}(b) and (d).}
\tablenotetext{c}{estimated at half the angle of the dark cusp in 171 {\AA} in Figure~\ref{evocut}(d).}
\tablenotetext{d}{via Equation~(\ref{inspeed}).}
\tablenotetext{e}{from Figure~\ref{allts}(c).}
\tablenotetext{f}{approximated as the mean of the HMI longitudinal magnetic strength above a noise level $\sim$ 10 G \citep{liuy2012} for the magenta circles in Figure~\ref{evocut}.}
\tablenotetext{g}{via Equation~(\ref{binflow}) and transformed to absolutes.}
\tablenotetext{h}{through the method in Section~\ref{dem}.}
\tablenotetext{i}{approximated as the diameter of the magenta circle in Figure~\ref{evocut}.}
\tablenotetext{j}{via Equation~(\ref{ne2}).}
\tablenotetext{k}{through Equation~(\ref{alfspeed}).}
\tablenotetext{l}{via Equation~(\ref{mach}) or (\ref{mach2}).}
\tablecomments{These estimates are made at $\sim$ 03:50 UT, just before the GOES 1-8 {\AA} flux peaks. The method for estimating the reconnection rate $M_A$ in the last column is described in Section~\ref{rate}.}
\end{deluxetable*}

\begin{figure*}
\begin{centering}
\includegraphics[scale=0.9]{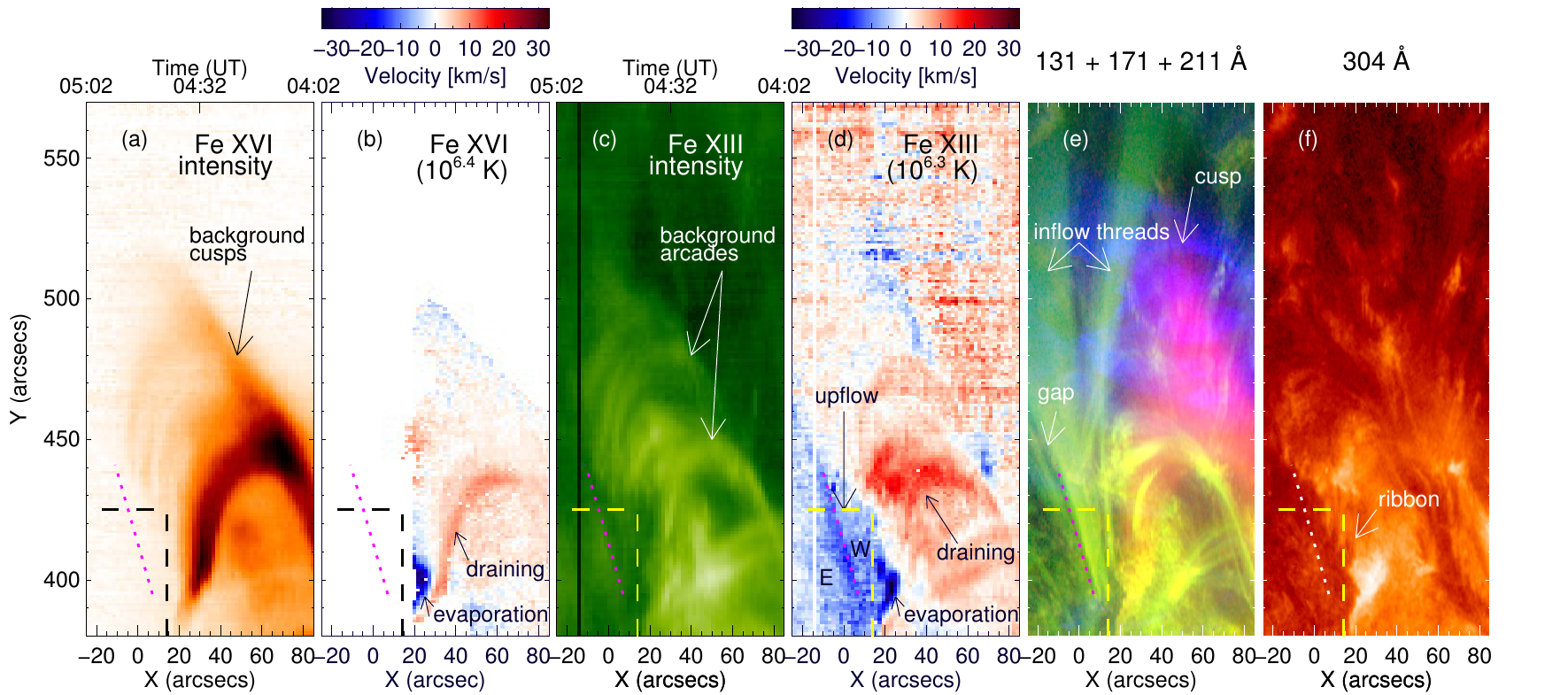}
\caption{\label{intvel}(a)-(b) Fe XVI 262.98 {\AA} intensity and Doppler velocity maps. (c)-(d) Fe XIII 202.04 {\AA} intensity and Doppler velocity maps ((Fe XII 195.12 {\AA} intensity and Doppler velocity maps are not shown here as they are similar to the ones of Fe XIII)). The sampling times of the EIS slit are added above (a) and (c). (e)-(f) synthesized AIA images simulating the EIS slit scanning mode for comparison. 131 {\AA} is red, 171 {\AA} green, and 211 {\AA} blue in (e). To align with EIS observations, they have not been rotated like in Figure~\ref{evocut}. The dashed line at the bottom left corner encloses the extended blueshift area in (d). The magenta dotted line is for the longitudinal velocity profiles in Figure~\ref{velden}.}
\end{centering}
\end{figure*}

\paragraph{Plasma downflows}
Figures~\ref{intvel}(a) and (c) show the Fe XVI and Fe XIII intensity maps from EIS, and Figures~\ref{intvel}(b) and (d) the corresponding line-of-sight velocity maps. For comparison, Figures~\ref{intvel}(e) and (f) are synthesized AIA ``raster'' images which simulate the EIS slit scanning mode, produced by combining narrow slices of AIA images at the EIS slit locations and times. Looptops and loop legs of the flare arcades (Figure~\ref{intvel}(b) or (d)) have redshifts of $\sim13~\rm km~s^{-1}$ indicating plasma draining, or loop contraction. We consider plasma draining to be the more likely explanation as the line-of-sight speed is much larger than the projected contraction speed $\sim1~\rm km~s^{-1}$ obtained from the hotter 94 {\AA} observations at that time (Figure~\ref{allts}(b)). An interpretation in terms of contraction is thus difficult to reconcile with the observed arcade geometry.

\paragraph{Plasma Upflows}\label{upflow}
We also have evidence of plasma upflows at the edge of the AR. The strong blueshift $\sim25~\rm km~s^{-1}$ at the eastern footpoint of the cusp (at $\rm (X, Y)\sim(25'', 400'')$ in Figures~\ref{intvel}(b) and (d)) could indicate chromospheric evaporation onto the reconnected cusp field (Figures~\ref{intvel}(a) and (e)). Just to its east is an extended blueshift area (enclosed by the yellow dashed line at the bottom left corner in Figure~\ref{intvel}(d)). This area can be divided into three parts, the strongest blueshift feature indicated by the magenta dotted line, the ``E'' region to the east, and the ``W'' region to the west. The ``W'' region possesses stronger blueshift than the ``E'' region. Note that the strongest blueshift feature in this area is well aligned with the gap with weak emission in the composite AIA image in Figure~\ref{intvel}(e), which boosts our confidence in the accuracy of the alignment between EIS and AIA.

\begin{figure}
\begin{centering}
\includegraphics[scale=0.45]{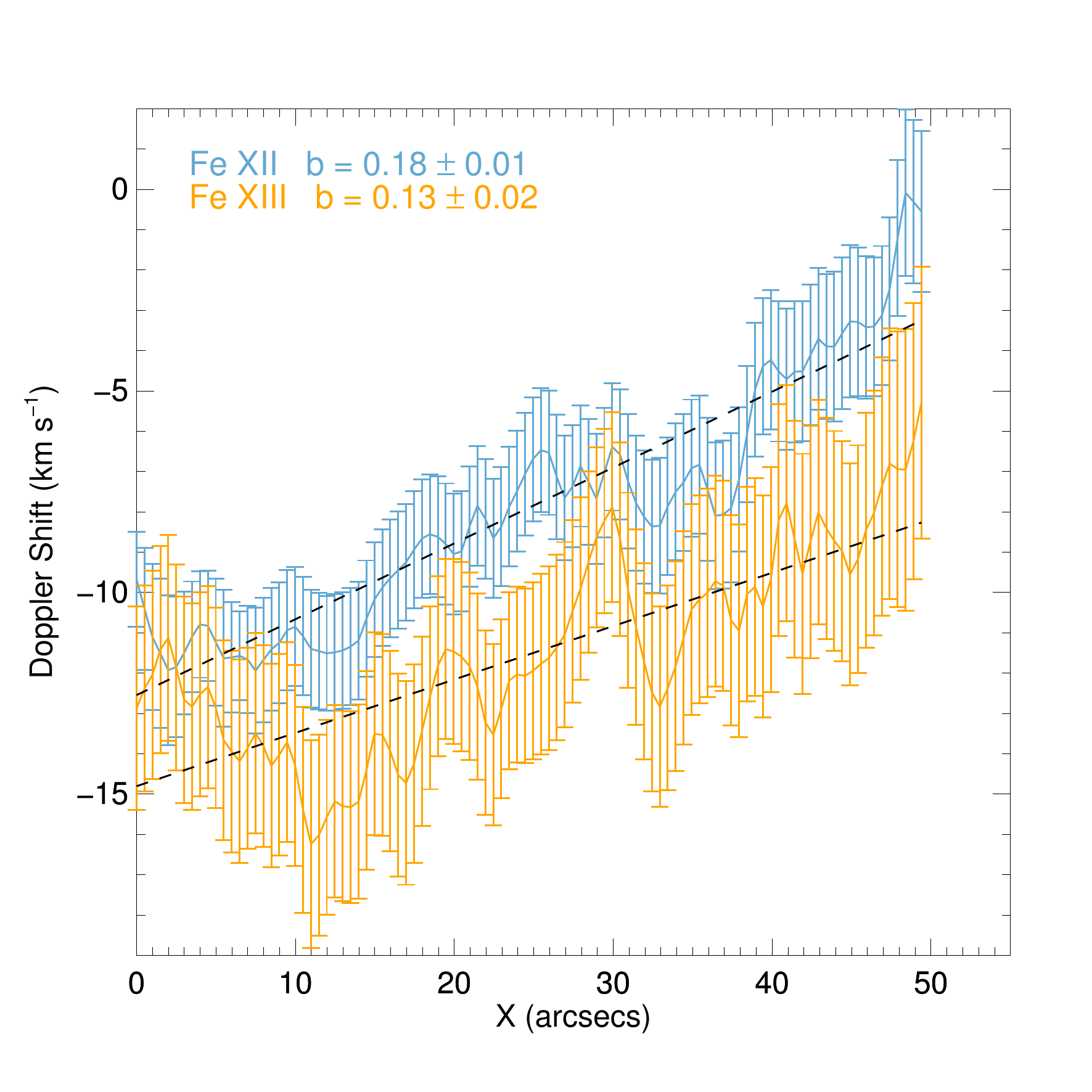}
\caption{\label{velden} Longitudinal velocity profiles for Fe XII and Fe XIII along the dotted line in Figure~\ref{intvel}. The origin of the X axis represents the bottom of the dotted line. The dashed line is the linear fit for each profile. ``b'' represents the slope of the fit and its $1-\sigma$ uncertainty. The uncertainty for the rest wavelength estimation is $\sim5~\rm km~s^{-1}$ for both lines, which would shift the entire profiles up or down.}
\end{centering}
\end{figure}

By comparing Figure~\ref{intvel}(d) with (e), it can be seen that the field corresponding to this extended blueshift area has not yet been reconnected in the main flare related to the bright cusp, so the blueshifts cannot be explained by the evaporation from the main flare. Nor can they be attributed to evaporation in the background, as the 304 {\AA} ribbon in Figure~\ref{intvel}(f) has not reached this area. For the strongest blueshift feature indicated by the magenta line, which is just to the east of the edge of the inflow threads, we can also exclude it being due to changing field inclination. If the line-of-sight velocity profiles along the dotted line, shown in Figure~\ref{velden}, were completely due to the inflow threads inclining towards us, we would expect a blueshift around zero at the footpoints and increasing with altitude. The observation in Figure~\ref{velden} contradicts this. Figure~\ref{velden} also excludes a loop siphon flow, in which the flows accelerate towards higher altitudes \citep{asc2005}. An easy way to interpret the blueshift along the dotted line is to invoke a plasma upflow along a field which inclines towards us. The same argument also applies to the ``W'' region. For the ``E'' region, it is difficult to argue as the velocity values are comparable to the rest wavelength uncertainty.

\begin{figure}
\begin{centering}
\includegraphics[scale=0.45]{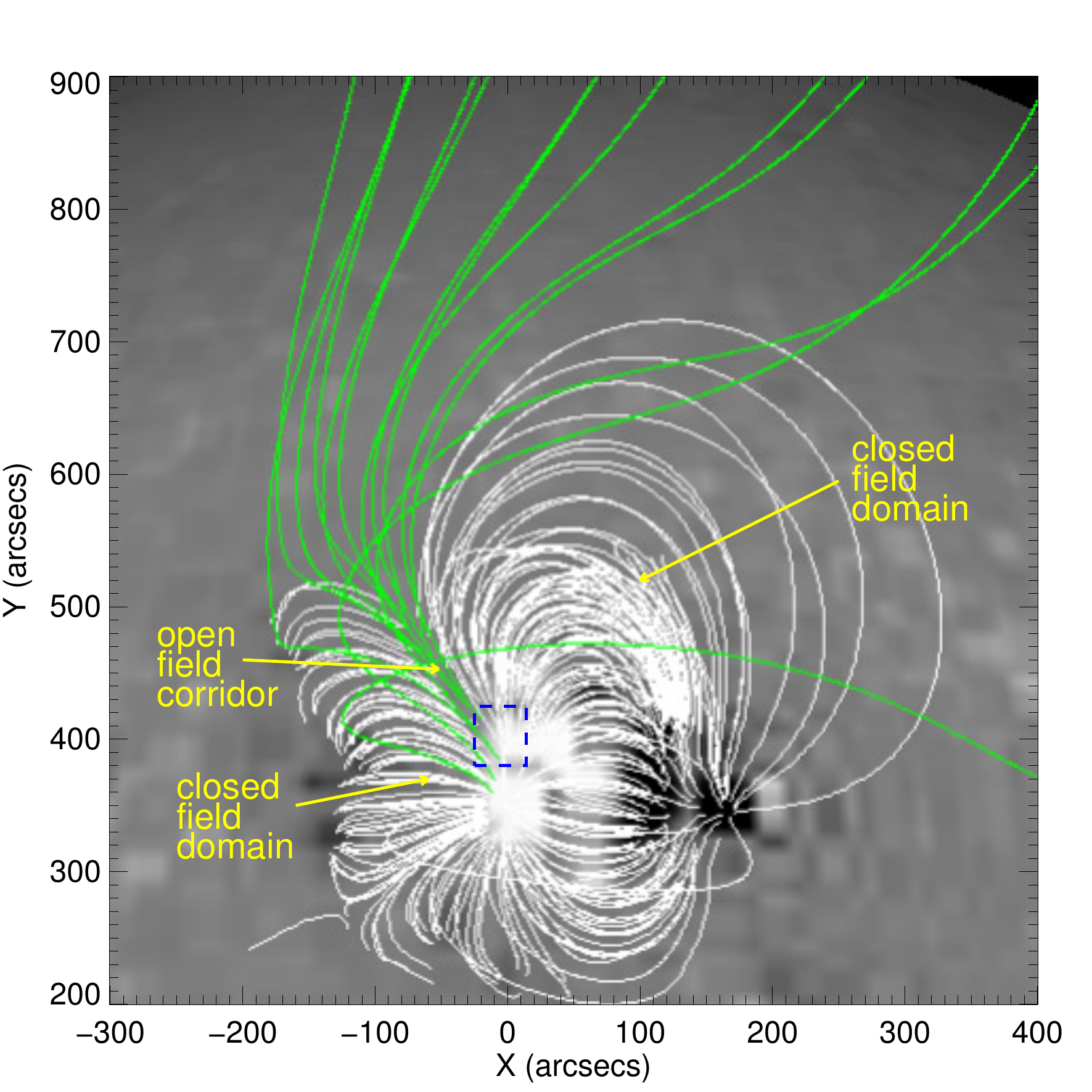}
\caption{\label{pfss}Model field at 00:04 UT just before the arcade eruption and the flare, derived from the PFSS package of Solarsoft, whose perspective has been rotated to 04:50 UT in order to compare with the extended blueshift region observed by EIS in Figure~\ref{intvel}(d). It shows a narrow open-field corridor between two closed-field domains. The open-field corridor extends northwards to a coronal hole. The blue dashed box shows the same region as the bottom left corner enclosed by the dashed line in Figure~\ref{intvel}.}
\end{centering}
\end{figure}

\subsection{Electron Density Estimate}\label{dem}
Differential emission measure (DEM) analysis can be used to estimate the electron density $n_e$ \citep{han2013}. The DEM is defined as $\xi(T)=n_e^2 dl/dT$ \citep{cra1976}, and integrating over $T$ results in the emission measure along the line-of-sight $\mathrm{EM}=\int \xi(t) dT=\int n_e^2 dl$. 

We calculate the emission measure during ($\mathrm{EM}_{fl}$ at 03:50 UT) and before ($\mathrm{EM}_{pre}$ at 03:00 UT) the inflow, 
using the regularization method of \cite{han2012} to recover $\xi(T)$ 
from the mean intensity in each of the 6 AIA wave bands (94, 131, 171, 193, 211, 335 {\AA}) with single exposures for both the eastern and western inflow regions (the two cyan boxes in Figure~\ref{evocut}(c) and (d)).
The temperature range used as input is $10^{5.5}$-$10^{6.6}$ K. The resulting DEM enhancement caused by the inflow concentrates between $10^{5.8}$ K and $10^{6.3}$ K, consistent with AIA observations, as the inflow threads can be most clearly seen in AIA 171 {\AA} which is more sensitive to this temperature range compared to other filters. However, the resulting EMs also contain a contribution from the background and foreground corona.
If we assume that (i) 
the background and foreground density outside the inflow threads does not change much during the event, 
and 
(ii) the density within the inflow region during the inflow (Figure~\ref{evocut}(d)) is much larger than before the inflow (Figure~\ref{evocut}(c)), then we can obtain the emission measure of the inflow region $\mathrm{EM}_{in}$ by taking
\begin{equation}
\mathrm{EM}_{in} = \mathrm{EM}_{fl} - \mathrm{EM}_{pre}.
\end{equation}
The electron density of the inflow region can then be estimated by:
\begin{equation}\label{ne2}
n_e = \sqrt{\mathrm{EM}_{in}/L},
\end{equation}
with $L$ being the line-of-sight thickness of the inflow region.

As $n_e\varpropto L^{-0.5}$ in Equation~(\ref{ne2}), the estimated density is not very sensitive to the choice of the thickness $L$. Thus we choose the diameter of the magenta circle in Figure~\ref{evocut} as an approximation of the thickness of the inflow threads, $L = 15\ \mathrm{arcsecs} \approx 1.1 \times 10^9$~cm. We then find $n_e \approx 2.1 \times 10^{8}$ cm$^{-3}$ and $\approx 2.0 \times 10^{8}$ cm$^{-3}$ for the eastern and western regions, respectively (Table~\ref{para}).

Assumption (i) above seems reasonable as no obvious events (except the inflow) happen during this period along the chosen boxes' line of sight. Assumption (ii) could be true, as firstly in the pre-inflow stage the two boxes are located within the dark void region (Figure~\ref{evocut}(c)), which means lack of emitting plasma, and secondly the void expansion may further evacuate the plasma there. And the obtained results above are consistent with EIS density diagnositics using Fe XIII 202.04 and 203.83 {\AA} pair ($\approx 1.5 \times 10^{8}$ cm$^{-3}$) around the same regions, though the EIS sampling time is after 04:00 UT (as can be seen in Figure~\ref{intvel}) and the reconnection site has already moved upwards.

\subsection{Magnetic Reconnection Rate}\label{rate}
The magnetic reconnection rate can be represented by the inflow Alfv\'{e}n Mach number
\begin{equation}\label{mach}
  M_{A}=V_{\rm in}/V_{A}
\end{equation}
where $V_{\rm in}$ is the inflow speed and $V_{A}$ the local Alfv\'{e}n speed. $V_{\rm in}$ can be estimated using
\begin{equation}\label{inspeed}
V_{\rm in}=V_{\rm patt}-V_{\rm xp}\tan{\theta}
\end{equation}
as in \cite{yok2001}, where $V_{\rm patt}$ is the apparent inflow speed obtained from the pattern of inflowing threads, $V_{\rm xp}$ the rising speed of the reconnection X-point, and $\theta$ the angle between the inflow threads and the rising direction of the X-point. This equation accounts for the rising motion of the reconnection site. The Alfv\'{e}n speed $V_{A}$ is
\begin{equation}\label{alfspeed}
 V_{A}=\frac{B_{\rm in}}{\sqrt{4\pi\rho}}\approx\frac{B_{\rm in}}{\sqrt{4\pi\mu{m_H}{n_e}}}
\end{equation}
in Gauss units, where $B_{\rm in}$ is the magnetic field strength in the inflow region, $\rho$ the mass density, $\mu$ the mean atomic weight ($\sim1.27$ for coronal abundances; \citeauthor{asc2005} \citeyear{asc2005}), $m_H$ the hydrogen mass, and $n_e$ the electron number density. To obtain $B_{\rm in}$, conservation of magnetic flux can be exploited \citep[e.g.][]{iso2002}, 
\begin{equation}\label{binflow}
B_{\rm in}V_{\rm in}=B_{\rm foot}V_{\rm foot}
\end{equation}
where $B_{\rm foot}$ is the vertical magnetic strength at the photosphere and $V_{\rm foot}$ the separation speed of flaring ribbons. As this AR is close to the solar disk center, HMI longitudinal magnetograms can be used as a good approximation of the vertical field. By combining Equations (\ref{mach}), (\ref{inspeed}), (\ref{alfspeed}) and (\ref{binflow}), the final equation for the reconnection rate is
\begin{equation}\label{mach2}
 M_A=\frac{(V_{\rm patt}-V_{\rm xp}\tan{\theta})^2}{B_{\rm foot}V_{\rm foot}}\sqrt{4\pi\mu{m_H}{n_e}}
\end{equation}
where the electron number density can be estimated as in Section~\ref{dem}, and other quantities are obtained as described in the notes to Table~\ref{para}. The resulting reconnection rates are 0.03 for both the eastern and western inflows.

\section{Discussion and Conclusions} \label{discussion}
We have reported the first comprehensive observations of reconnection flows on the solar disk. Threads or strands of plasma accelerate and later decelerate towards a presumed reconnection site, below which a well-defined hot cusp forms, anchored at the threads' endpoints. Individual cusp loops shrink and cool as the brightest portion of the cusp ascends. 
The magnetic reconnection rates around the GOES flux peak are 0.03 for both the eastern and western inflows, consistent with fast reconnection, and in the range of previous studies \citep{yok2001,lin2005,nar2006,bem2010,sav2012,su2013,sunj2015,zhu2016}. The reconnection is quite symmetric in this case.
According to Equation~\ref{mach2}, if $V_{xp}$ and $\theta$ are good observational estimates, the reconnection rate estimated is most sensitive to $V_{patt}$, only the transverse component of the real inflow velocity. For a rough estimation of the lower limit of the reconnection rate, we double $B_{foot}$, $V_{foot}$ and reduce $n_e$ by a factor 10, giving reconnection rates of around 0.003 for both the eastern and western inflows, which are still in the fast reconnection regime.

There is no emission from the presumed reconnection site; it may be too short or thin, or at the wrong temperature to be detected by the instruments used. We note that the upper part of the dark cusp highlighted in Figure~\ref{evocut}(d) is dark in all AIA wavelengths, implying that it has a very low density, or temperature above the $\sim 10$ MK at which the AIA 131~\AA~filter peaks and where the cusp is clearest.

As argued in Section~\ref{allflows}, possibilities like evaporation from the main flare, field inclining and a siphon flow, could not be the reasons for the blueshifts along the dotted line and in the ``W'' region in Figure~\ref{intvel}(d). Plasma upflows along field which inclines towards us could be an explanation for these blueshifts. Blueshift features are found to be ubiquitous at the edge of ARs from EIS observations even in non-flaring regions, persisting from hours to days in areas of weak emission and low density, and possessing velocities around tens of $\rm~km~s^{-1}$, faster in hotter lines \citep[e.g.,][]{sak2007,del2008,dos2008,har2008,bak2009,dem2013,bro2015,del2008,dos2008}. They are interpreted as uplfows by some authors and considered to be a possible source of the slow solar wind in the heliosphere, but the real origin of these blueshift features is still controversial \citep[][and references therein]{abb2016}.

We here propose a distinction between two upflow components associated with the blueshift features observed in this event. The strongest blueshift in Figure~\ref{intvel}(d) is well aligned with the gap with weak AIA emission in Figure~\ref{intvel}(e) which may imply open field short of emitting plasma, while the ``W'' region evidently corresponds to the large-scale closed loops which are the inflow threads or the legs of the arcade loops erupting outwards\footnote{The upper part of the inflow threads could be contaminated by the background arcades which have draining plasma.} in Figure~\ref{intvel}(e). The potential-field source-surface (PFSS) model just before the flare in Figure~\ref{pfss} provides supportive evidence. It well reflects the pre-eruption structure seen in Figure~\ref{evocut}(a), and shows that the extended blueshift area in Figure~\ref{intvel}(d) consists of a mix of open and closed fields. Two closed-field domains are separated by a very narrow open-field corridor, which matches with the structure in the extended blueshift area in Figure~\ref{intvel}(d) with the strongest blueshift feature indicated by the dotted line seperating the ``E'' and ``W'' regions apart. Thus it seems that plasma upflows occur along both open field and large-scale closed loops. The argument above helps solve a long-standing problem that whether the blueshift-related upflows at the AR boundary are associated with open or large-scale closed field \citep{sak2007,har2008,bak2009,del2011,bou2012,bro2015,edw2016}.

As the blueshift levels of the feature indicated by the dotted line in Figure~\ref{intvel}(d) and the ``W'' region are quite different (collimated and stronger along the dotted line), different mechanisms may be responsible for the associated upflows. For the upflow in the ``W'' region, expansion of related large-scale closed loops \citep{har2008} could be an explanation. When the flux tube of the arcade expands outwards, the plasma within would diffuse upwards because of pressure imbalance. We here suggest that the diffusion may not be adiabatic, and it could be a diffusion with a source at the bottom of the corona. The dilution of the plasma within the flux tube as it expands can result in ``depressurization'' \citep[][see the third row of its Figure~9]{ree2010}, which would be able to induce a plasma upflow from the coronal base along the legs of the expanding arcade. A vivid analogy of this ``depressurization'' is the water in a tube being pumped out by rapidly pulling a plunger. As the expanding arcade's legs are the inflow threads here, the upflow due to the ``depressurization'' may serve as a way to increase the plasma density advected into the reconnection region or other acceleration regions (e.g., the slow-mode shock), which could help relax the ``electron number problem'' \citep{bro1977,fle2008} to some extent. \textit{Fermi} Gamma-ray Burst Monitor \citep[GBM;][]{mee2009} observations barely show any hard X-ray emission from this flare (unfortunately also no observations from the Reuven Ramaty High Energy Solar Spectroscopic Imager for this event), implying a very weak requirement for the electron flux. For a major flare, the eruption and arcade expansion could be more violent, possibly with a faster upflow and increased electron supply.

\citet{ant2011} shows that a narrow open-field corridor maps to separatrices and quasi-separatrix layers (QSLs) in the heliosphere where the magnetic connectivities change dramatically, and they are the natural region for interchange reconnection between open and closed field to take place \citep{fis1999,fis2003}. Thus the uplfow associated with the open field here in Figure~\ref{pfss} could be created by reconnection between the open field and the two closed domains nearby, which transports plasma from the closed field to the open one. Comparing Figure~\ref{evocut}(a) with (f), it can be seen that the intensity of the eastern closed domain has a significant decrease during the evolution while the large-scale loops nearby to the west become more intense, which could mean that an interaction happens between the eastern closed domain and the narrow open field corridor. The main flare or the arcade eruption observed in the western domain may facilitate or impede the dynamics.

A characteristic inclination angle of the open field in Figure~\ref{pfss} towards us can be obtained from the PFSS model to be $\sim45$\textdegree. Figure~\ref{intvel}(d) (and also Figure~\ref{intvel}(b)) provides the characteristic values of the longitudinal velocities of the blueshift feature indicated by the dotted line, the ``W'' region, the evaporation feature, and the plasma draining, to be $\sim10~\rm km~s^{-1}$, $\sim5~\rm km~s^{-1}$, $\sim25~\rm km~s^{-1}$, and $\sim13~\rm km~s^{-1}$, separately. If we assume that all the fields related to the above features incline towards us with roughly the same angle $\sim45$\textdegree~ as the open field does, the total speeds of the associated plasma flows travelling along these fields can be estimated to be $\sim14~\rm km~s^{-1}$, $\sim7~\rm km~s^{-1}$, $\sim35~\rm km~s^{-1}$, and $\sim18~\rm km~s^{-1}$, respectively. They are all subsonic as the sound speed for a plasma with a temperature $T_e\sim2.0~\rm MK$ or $\sim2.5~\rm MK$ (for Fe XIII $\sim10^{6.3}~\rm K$ and Fe XVI $\sim10^{6.4}~\rm K$, respectively) is $c_s=147\sqrt{{T_e}/{1~\rm MK}}\sim208~\rm km~s^{-1}$ or $\sim232~\rm km~s^{-1}$ \citep{asc2005}. The upflow speeds from a few to tens of $\rm km~s^{-1}$ at the edge of the AR are consistent with previous EIS observations \citep{del2008}. The evaporation speed $\sim35~\rm km~s^{-1}$ is similar to the results obtained by \citet{mil2009} also for a C class flare at this temperature range. The plasma draining speed $\sim18~\rm km~s^{-1}$ is also comparable to previous results derived from EIS spectroscopy \citep{del2008,syn2012} though they measured at the quiet stage of the AR evolution. The plasma draining at these spectral lines may reflect the warm counterpart of the cold coronal rain \citep[e.g.,][]{sch2001,kam2011,vas2015} observed later in 304 {\AA}.

In addition, if we take the field inclination into account when calculating the reconnection rate, this will slightly change the values of $V_{xp}$ and $\theta$ in Table~\ref{para}, but the final reconnection rate around the GOES flux peak will still be rounded to 0.03 for both the inflow regions and in agreement with fast reconnection.

Together with \citet{li2017}, this work reveals the 2D and 3D aspects of this event.
The wealth of diagnostic information on the flows and plasma properties around the reconnection region and at the periphery of the AR can be further used to explore the energetics of the reconnection process and the detailed dynamics of flow evolution, while the availability of HMI vector magnetograms means that the magnetic evolution and plasma flows can be investigated in more detail using magnetic field extrapolations and magnetohydrodynamic simulations.

\acknowledgments
The authors thank the referees for very helpful comments, which significantly improve the quality of the manuscript, and also Hugh Hudson for discussion and comments. L.\,F., P.\,J.\,A.\,S., and N.\,L.\,S.\,J. acknowledge support from STFC Consolidated Grant ST/L000741/1. The research leading to these results has received funding from the European Community's Seventh Framework Programme (FP7/2007-2013) under grant agreement No. 606862 (F-CHROMA). P.\,J.\,W is supported by an EPSRC/Royal Society Fellowship Engagement Award (EP/M00371X/1). I.\,G.\,H. acknowledges support of a Royal Society University Research Fellowship. The authors are grateful to NASA/SDO, AIA, HMI, Hinode/EIS, and GOES science teams for data access.

{\it Facilities:} \facility{AIA (SDO)}, \facility{HMI (SDO)}, \facility{EIS (Hinode)}, \facility{GOES}.


\end{document}